\documentclass[preprint,12pt,authoryear]{elsarticle}
 
\usepackage{amsmath,amsfonts,amssymb}
\usepackage{graphicx}

\usepackage{aas_macros}
\usepackage{natbib}
\usepackage{url}
\usepackage{amssymb}
\usepackage{amsmath}
\usepackage{amsfonts}

\journal{Nuclear Instr. and Methods in Phys. Res. A}

\begin{document}

\begin{frontmatter}



\title{Correlation methods for the analysis of X-ray polarimetric signals}

\author[label1,label2]{E. Massaro} 
\author[label1]{S. Fabiani}
\author[label3]{R. Campana}
\author[label1]{E. Costa}
\author[label1]{E. Del Monte}
\author[label1]{F. Muleri}
\author[label1]{P. Soffitta}%\author[label1]{~~~~~~~P. Soffitta}

\address[label1]{IAPS-INAF, Roma, Italy}
\address[label2]{In Unam Sapientiam, Roma, Italy}
\address[label3]{IASF-INAF, Bologna, Italy} 

\begin{abstract}
X-ray polarimetric measurements are based on studying the distribution of
the directions of scattered photons or photoelectrons and on the search of
a sinusoidal modulation with a period of $\pi$.
We developed two tools for investigating these angular distributions based
on the correlations between counts in phase bins separated by fixed
phase distances.
In one case we use the correlation between data separated by half of the
bin number (one period) which is expected to give a linear pattern.
In the other case, the scatter plot obtained by shifting by 1/8 of the bin number
(1/4 of period) transforms the sinusoid in a circular pattern whose radius is
equal to the amplitude of the modulation.
For unpolarized radiation these plots are reduced to a random point distribution
centred at the mean count level.
This new methods provide direct visual and simple statistical tools for
evaluating the quality of polarization measurements and for estimating
the polarization parameters.
Furthermore they are useful for investigating distortions due to systematic
effects.
\end{abstract}


\end{frontmatter}



%Roma, version 1.6  2017 Feb. 12

\section{Introduction}

Measurements of the linear polarization in the X-ray band are based on the detection of
anisotropies either in the angular distribution of scattered photons or in the initial 
direction of photoelectrons (for a comprehensive review see the book by \cite{Fabiani2014a}).
The latter technique appears now more promising with the development of the Gas Pixel 
Detector (hereafter GPD) \citep{Costa2001, Bellazzini2006, Bellazzini2007} that will 
be the focal plane instrument of the first polarimetric mission after the pioneering  
age of OSO 8.
   \begin{figure} [ht]
   \begin{center}
   \begin{tabular}{c} %% tabular useful for creating an array of images 
   \includegraphics[height=8cm]{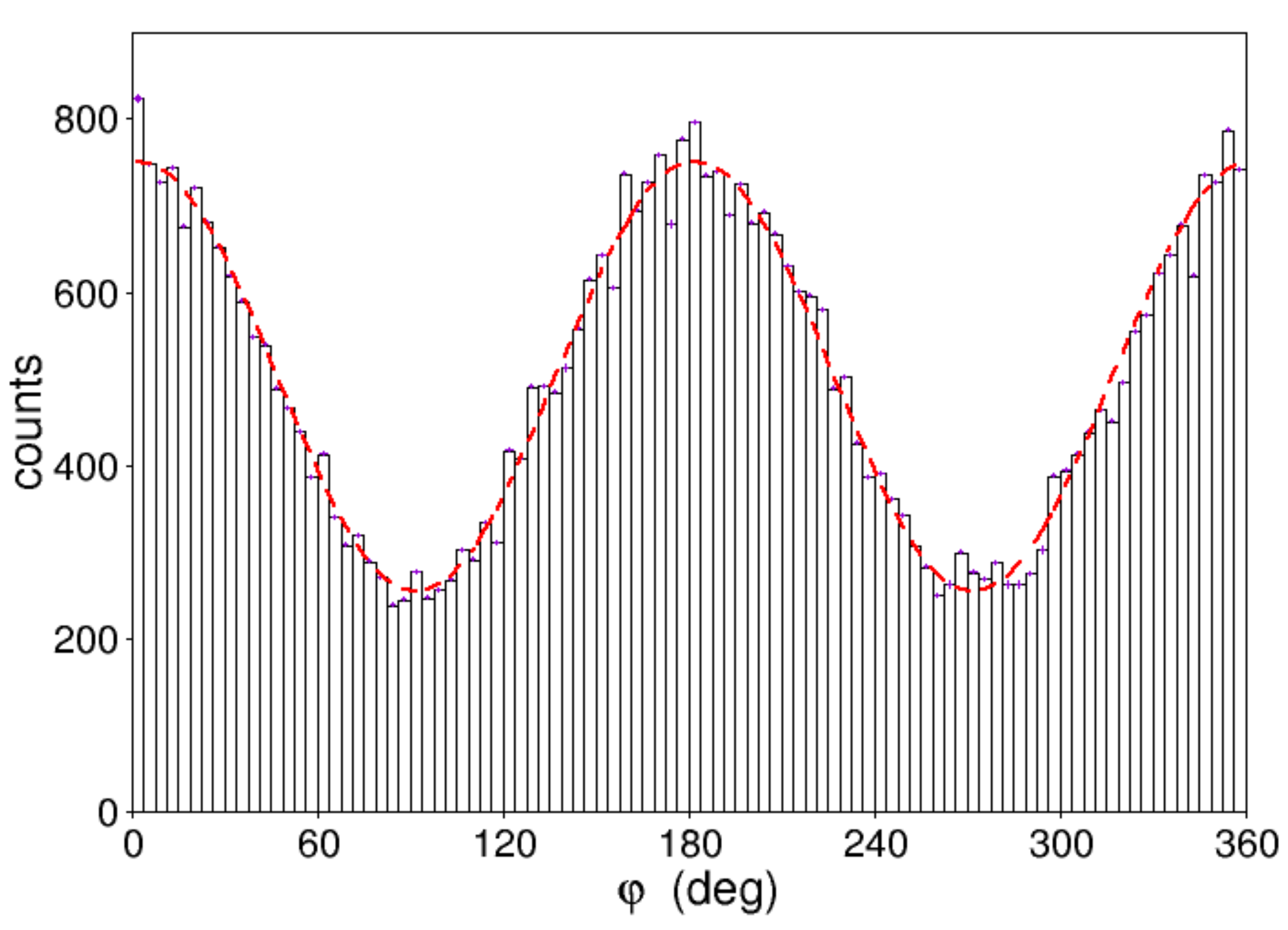}
   \end{tabular}
   \end{center}
   \caption[example] 
%>>>> use \label inside caption to get Fig. number with \ref{}
   { \label{fig:hist} 
Histogram of the angular distribution of the initial direction
of electrons due to a polarized X-ray flux.
The dashed curve is the best fit modulation.}
   \end{figure} 
Indeed IXPE (Imaging X-ray Polarimetry Explorer, \cite{Weisskopf2016}) has been selected 
as the next SMEX NASA mission for a flight in late 2020. 
With three mirrors and three GPDs it will perform spectral-temporal-angular resolved 
polarimetry as a break-through measurement in astrophysics, allowing the opening of a 
new window on the high-energy sky. 
Meanwhile XIPE (X-ray Imaging Polarimetry Explorer) \citep{Soffitta2016}, has 
completed its phase A study in the framework of the $4^{th}$ ESA call for a medium size 
mission.

In the present paper we show some simple correlation methods useful for the detection and for
the parameter estimates of a polarized signal. 
These methods are based on the phase distribution histograms of photons (or electrons) 
and do not require measurements of the Stokes' parameters; for this reason they have the 
advantage of simplifying the statistical tools used for evaluating the significance of 
the polarimetric estimates \citep{Massaro2016}. Our new method is therefore complemetsry to the classical Stokes' analysis.

This topic has been previously studied by different authors. The relation between sensitivity 
and significance of a polarization measurement \citep{Weisskopf2010, Elsner2012} has been 
revisited by means of simulation \citep{Strohmayer2013}.
The use of Stokes' parameters for X-ray polarimetry has been explored by \citet{Kislat2015} 
while the possibility to extend the standard tools of X-ray astronomy (XSPEC) to polarization 
by using Stokes parameter has been studied very recently by \citet{Strohmayer2017}. 
Other authors, instead, investigated the possibility of using Bayesian methods to derive the 
polarization angle and degree, which may be interesting when low signal-to-noise ratio are expected 
\citep{Vaillancourt2006, Maier2014}. 

An important issue in polarization measurements is the possibility to have
a noise distribution differing from the Poissonian one because of intrinsic
effects in the detection technique. For instance, in a photoelectric polarimeter
as the GPD the uncertainty on the initial direction of the
photoelectron can be influenced  by the hexagonal geometrical pattern of pixel plane (expecially for tracks comprising a very small number of pixels) producing a not uniform spread of angular directions in the phase histograms \citep{Muleri2010}.

Thus the assumption that the noise on the angular distribution is white and
purely Poissonian cannot be always verified.
Our correlation methods are not based on this assumption and provide direct
visual tools for evaluating the quality of polarization measurements and for
investigating distortions due to systematic effects.

\section{Polarization data}
\label{sec:poldata}

For each detected photon the information on its linear polarization is given by the 
azimuthal angle $\varphi$ measured from the direction of a scattered photon (for scattering 
polarimeters) or from the initial direction of the photoelectron trajectory (for 
photoelectric polarimeters).
The detection of a polarization is therefore obtained from the angular distribution of the
$\varphi$ values of a number $N$ of observed photons and particularly from the amplitude 
of a sinusoidal modulation with a period of $\pi$ (or 180$^\circ$), due to the differential 
cross section of the physical process involved (Compton scattering and photoelectric absoption, 
respectively).
In the following we will refer to this histogram as the data set $\{n_k | k = 1,..., M\}$, 
with a total bin number $M$ and a number of events in the $k$-th bin indicated by $n_k$.
One can write the expected modulation curve as

 \begin{equation}
 n(\varphi_k) =  A~ \sin(2(\varphi_k + \psi))  + \langle n \rangle ~~~~, \label{eq:modulation2}
 \end{equation}

\noindent
where $\varphi_k = (2\pi/M)(k-1/2)$ is the phase of the bin centre, $A$ the amplitude 
of the modulation and $\psi$ is the polarization angle; for a total number of photons 
$N$ one obviously has
   \begin{equation}
\langle n \rangle = N/M  ~~~~~.
   \end{equation}
In Fig.~\ref{fig:hist} the histogram of a modulation curve of experimental data for a totally polarized source is  shown.
The polarization degree $p$ of the incoming signal is then evaluated by

 \begin{equation}
 p = \frac{1}{\mu} \frac{A}{\langle n \rangle}   ~~~~~,
 \label{eq:poldegree}
 \end{equation}

\noindent
where $\mu$ is the instrumental modulation factor, i.e. the amplitude resulting from
a totally polarized input radiation.
The noise is due to the statistical fluctuations of counts and to possible instrumental
effects.  

As it will be clear in the following it is advantageous to consider $M$ as a multiple of 8 
(in the following examples we will consider bin numbers with this property).

\section{The method of the Linear Correlation plot}

\begin{figure} [ht]
   \begin{center}
   \begin{tabular}{c} %% tabular useful for creating an array of images 
   \includegraphics[height=6.5cm]{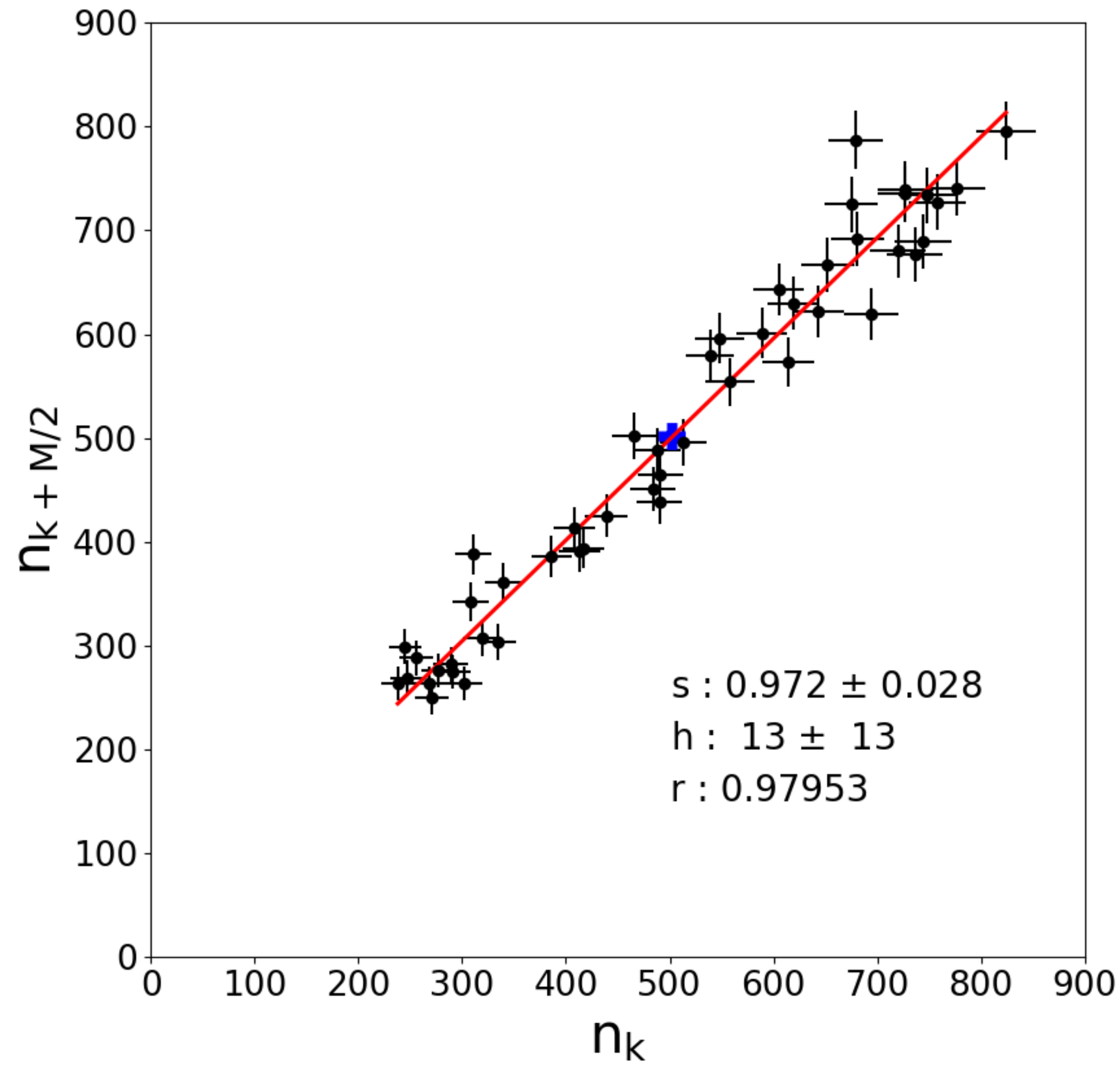} 
   \includegraphics[height=6.5cm]{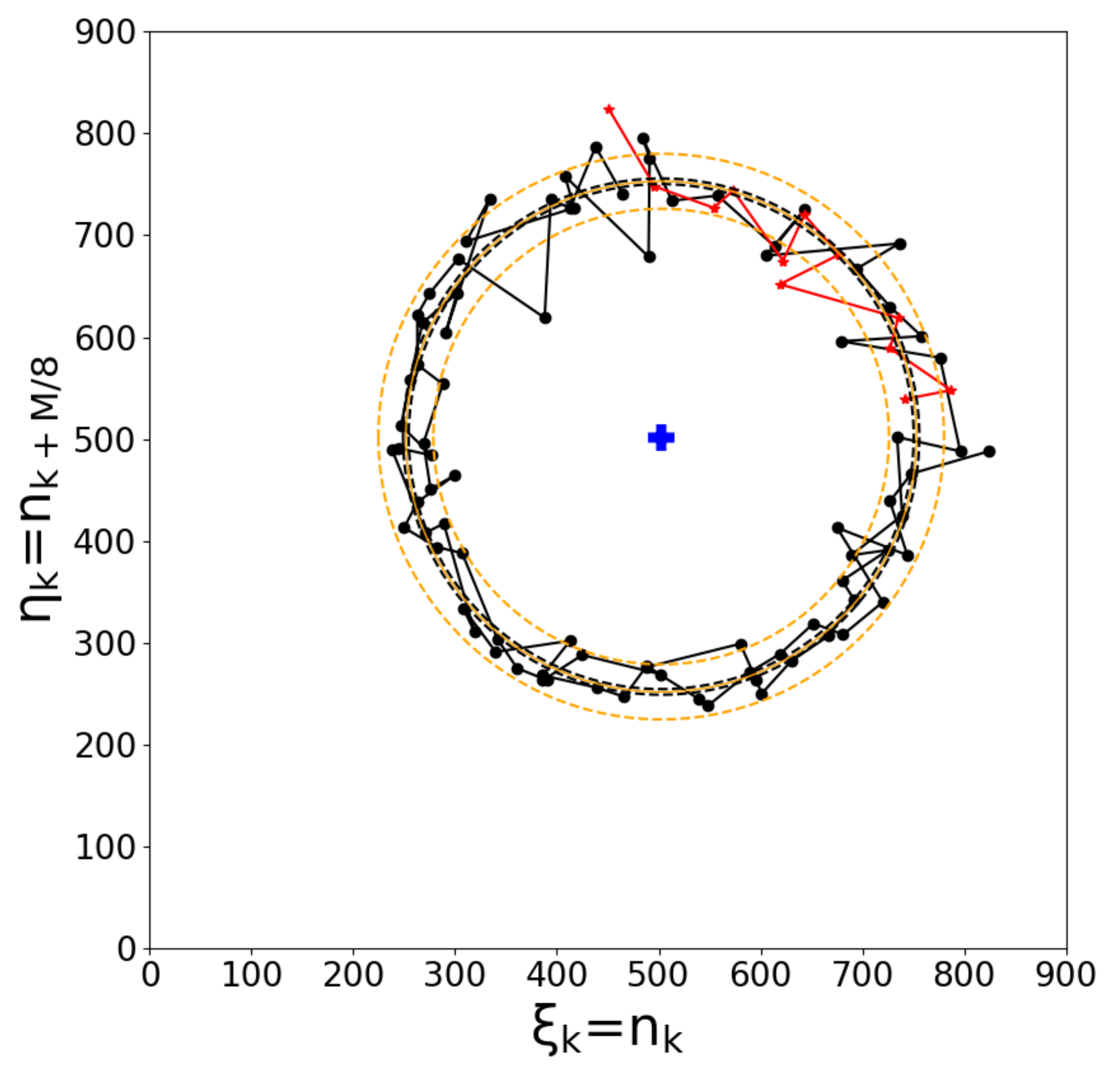}
   \end{tabular}
   \end{center}
   \caption[example] 
%>>>> use \label inside caption to get Fig. number with \ref{}
   { \label{fig:plotscirclelinear} 
Left panel: scatter plot exhibiting the linear correlation between the first and 
second sections of a 96 bin angular distribution in Fig.~\ref{fig:hist}. Errors are evaluated assuming a Poisson distribution of counts in the histogram bins.
The solid line is the linear best fit whose slope $s$ and intercept $h$ are given 
in the bottom of the figure with the correlation coefficient $r$. The fit is performed with the orthogonal distance regression method implemented in the python package \textit{scipy.odr} based on the ODRPACK FORTRAN-77 library (\url{https://docs.scipy.org/doc/external/odrpack_guide.pdf}).
The point representing the means of two sections' values is the bold cross.\\
Right panel: scatter plot of the same data exhibiting the circular pattern due to the 
shift of $M/8$ data shift in a 96 bin angular distribution.
Star points connected by the red line are those of the circular completion. The solid line circle has a radius equal to the average radial distance from the centre marked by the cross. The blue dashed lines closer to solid line circle show the uncertainty on the average radius, while the larger orange dashed lines represent the standard deviation of the distribution of radii. 
}
   \end{figure}

We create a scatter plot with $M/2$ points having coordinates 
$(n_k, n_{k+M/2})$ ($1 \leq k \leq M/2$).
If a polarization modulated signal is present, these point will be distributed along a 
straight segment with an inclination of 45$^\circ$ and with a length equal to the 
double of the modulation amplitude.
The linear correlation coefficient $r$ can provide useful information for estimating the statistical $S/N$ ratio. The significance of the polarization measurement is 
obtained from the likelihood to have a corresponding $r$ with $M/2$ degrees of freedom.
The linear plot for the same data set in Fig.~\ref{fig:hist} is given in 
Fig.~\ref{fig:plotscirclelinear} (left panel), where the values of the linear correlation 
coefficient and of the best fit line parameters are also reported toghether with the statistical uncertainties:
\begin{equation}
n_{k+M/2} = h + s~n_k
\end{equation}
\noindent
are also reported.

For a modulation as the one of Eq.~\ref{eq:modulation2} the expected values of the slope 
$s$ is 1 and of the constant $h$ is zero and it is actually found within 1 standard deviation as shown in Fig.~\ref{fig:plotscirclelinear} (left panel).
In the case of a non polarized radiation we expect no correlation between the 
phase bins and in the plot the corresponding points will be randomly distributed 
around the mean value, with a linear correlation coefficient close to zero.

It is well known from the correlation theory that $r^2$ is the fraction of the 
variance {\it explained} by the linear regression and that $1 - r^2$ is proportional to the 
{\it residual} variance due to the noise; thus the simplest way to evaluate 
the $S/N$ ratio of the polarization is given by $\sqrt{r^2/(1 - r^2)}$.
We stress that this simple formula is independent of any assumption of the nature 
of the noise and particularly if it is only due to the Poisson statistics or 
to the occurrence of possible systematic effects. In particular values of $s$ and/or $h$ non consistent with 1 and zero, respectivelly, likely indicate a systematic deviation.

A linear correlation with unit slope results not only from a sinusoidal pattern
like that of Eq.~\ref{eq:modulation2}, but also from any distribution with a period 
$M/2$, and thus this plot is not useful for detecting a polarized signal itself, 
but only for estimating the $S/N$ ratio and the occurrence of some systematic effects.
These in fact can introduce changes in the amplitude of the angular distribution
resulting in a value of the slope $s$ different from unity.
For instance, systematic deviations affecting an odd harmonic, like the 3rd one,
which are anticorrelated and therefore the expected value is $s = -1$, would 
produce a decrease of the slope (see Sect.~\ref{systematics}).
%\clearpage
%\clearpage

\section{The method of the Circular Plot}

A useful tool for measuring the polarization parameters and for a quick inspection
of data is based on the scatter plot of the histogram of the angular distribution
of events in the $(\xi , \eta)$ space, where these quantities are the histogram 
bin counts with the latter one shifted by $M/8$ bins.
Consider the distribution of $M$ points having the coordinates 
$(\xi_k = n_k, \eta_k = n_{k+M/8})$,~ ($1 \leq k \leq 7 M/8$),
completed with the $M/8$ points $(\xi_k = n_k, \eta_k = n_{k-7 M/8})$,~ 
($7 M/8+1 \leq k \leq M$).
If the counts are distributed with a $\sin(2\varphi)$ modulation the phase difference 
of $M/8$ corresponds to 1/4 of the period and the points in the 
$(\xi, \eta)$ space will be distributed in a circular pattern 
having the radius equal to the modulation amplitude.
Thus the problem of detecting a polarization is then reduced to that of 
finding such a circular distribution of data.

\subsection{Measure of the polarization degree and angle}

For each point of the plot we can compute the corresponding radius $R_k$  
   \begin{equation}
R_k = \sqrt{(n_k - \langle n \rangle)^2 + (n_{k+M/8} - \langle n \rangle)^2} = \sqrt{(\xi_k - N/M)^2 + (\eta_k - N/M)^2}
   \end{equation}

The distribution of $R_k$ against the phase bin number angle or the bin centre angle 
$\varphi_k$ is expected to be consistent with a constant and the ratio between the mean value of the radius 
$\langle R \rangle$ and its standard deviation $\sigma_R$ is an estimator of the $S/N$ ratio.
In Fig.~\ref{fig:plotscirclelinear}  (right panel) it is reported the circular polarization 
plot for the same data set of the left panel. The mean radius is $\langle R \rangle = (249.20 \pm 2.8)$.
The amplitude found by the usual method of a $\sin(2(\varphi - \psi))$ 
least-squares fitting results $A = (247.5 \pm 3.6)$, compatible with the previous estimate within 1$\sigma$.

Note that in the circular plot each value $n_k$ is used two times: first for computing 
$\eta$ and then for $\xi$; thus the $M$ values distributed in the two circles are not independent. 
For instance, a large positive fluctuation in one of the $\{n\}$ values would correspond 
to two values of $R_k$ in excess, the former one in the $\eta$ component and, after $M/8$ 
points, in the $\xi$ component.
According to Eq.~\ref{eq:poldegree}, the polarization degree $p$ is given by  

   \begin{equation}
p = \langle R \rangle/ (\mu~\langle n \rangle)  = \frac{\sum_k R_k}{\mu N} ~~~~~~~ \label{eq.polarizationR}.
   \end{equation}

\begin{figure} [ht]
   \begin{center}
    \begin{tabular}{c} %% tabular useful for creating an array of images 
   \includegraphics[height=6cm,angle=0]{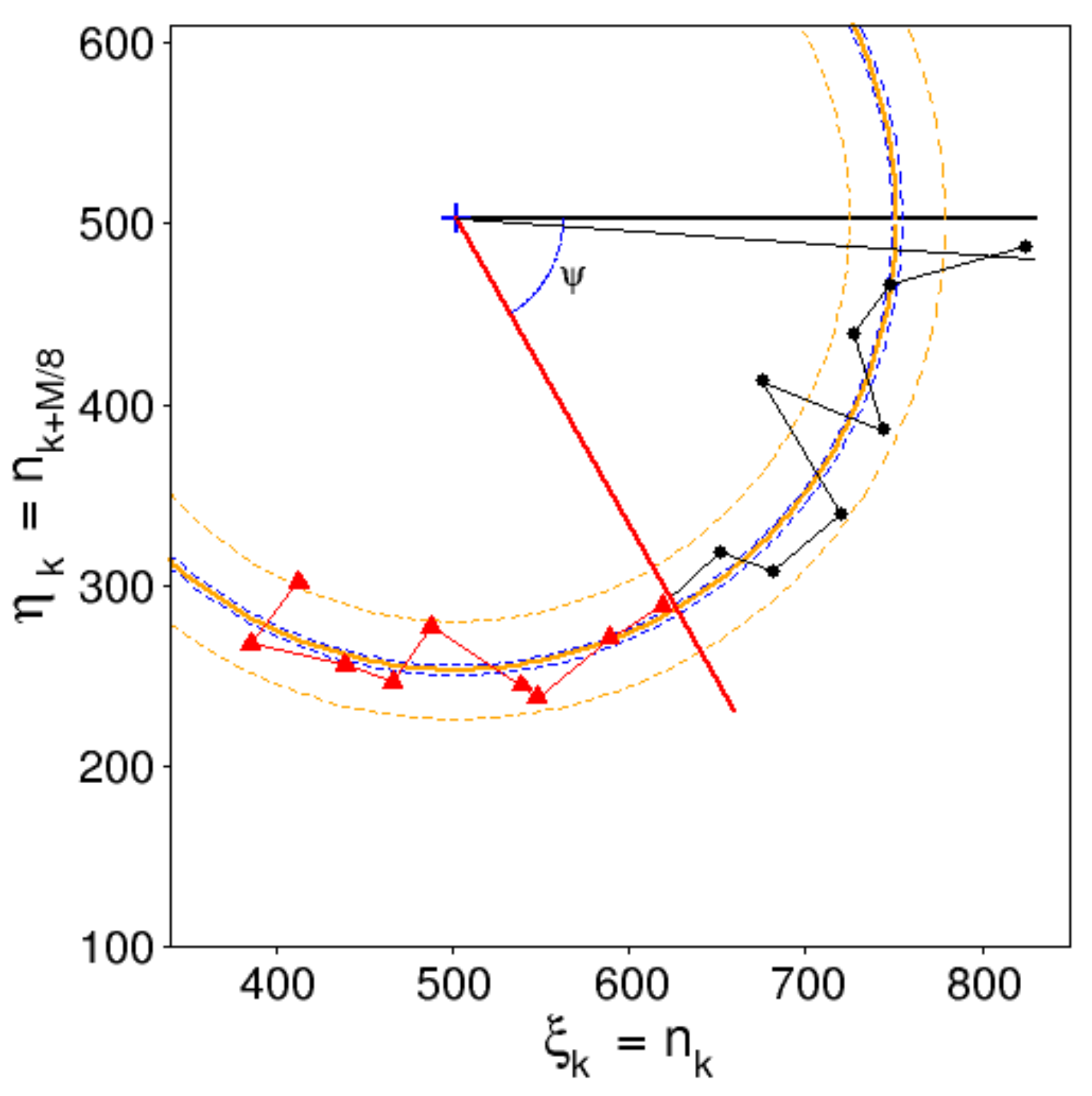}
   \includegraphics[height=6cm,angle=0]{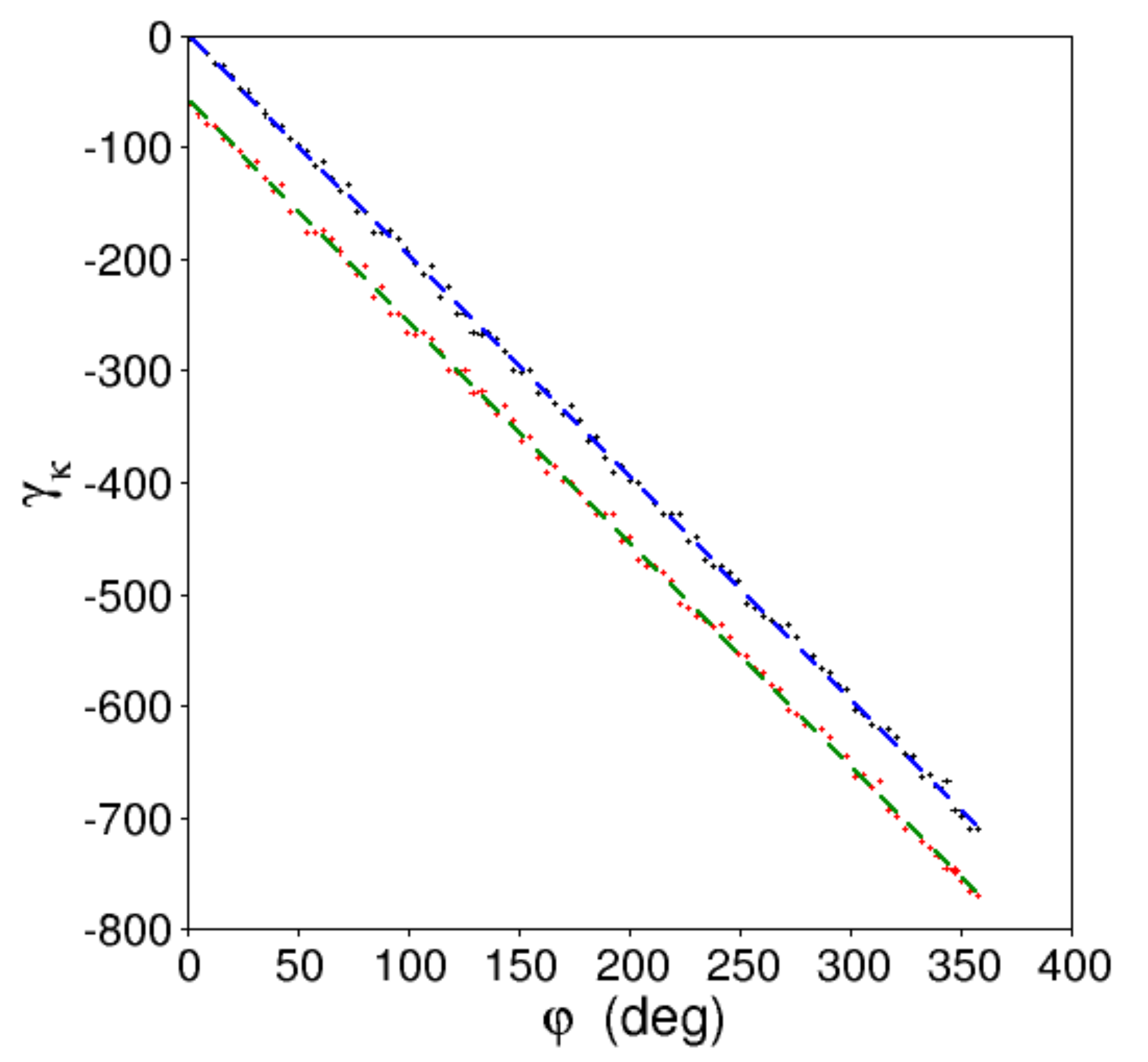}  
   \end{tabular}
  \end{center}
   \caption[example] 
%>>>> use \label inside caption to get Fig. number with \ref{}
   { \label{fig:anglin} 
Left panel: the initial part of the annular distribution of the original data and 
with a negative phase shif of 60$^{\circ}$. 
The starting point of the circular distribution is sensitive to the phase shift and 
therefore to the polarization direction.
Right panel: the linear relation between the angles $\gamma_k$, estimated from the 
circular plot in the lower panel of Fig.~\ref{fig:plotscirclelinear}, and the expected 
values from the bin phases $\varphi_k$.
Original data are the filled circles points and the dashed line is the best fit.
The lower data set (filled squared points) and their best fit (solid line)
are obtained from the same histogram applying a backward shift of 8 bins (equal to 
$30^{\circ}$).  
The slope best fit value is $-1.988 \pm 0.005$ and the the intercept values are 
$2^{\circ}.01 \pm 1^{\circ}.14$ and $-57^{\circ}.3 \pm 1^{\circ}.15$.}
   \end{figure}

Since the peaks of the modulation curve correspond to the maximum value of the $n_k$ coordinate with respect to the center of the circular correlation plot, the information on the polarization angle $\psi$ is also in the circular distribution. The polarization angle corresponds to the angle of the initial point of the data set with respect to the positive direction of the $\xi_k=n_k$ axis. By rotating the polarization angle the starting point of the circular correlation plot rotates. 
This is shown in left panel of Fig.~\ref{fig:anglin} where the same data of 
Fig.~\ref{fig:plotscirclelinear} are plotted (black dots) together with the points 
obtained by a negative phase shifting of 60$^{\circ}$ of the same histogram (red points). 
We can thus correlate the angles $\gamma_k$ of segments from the centre of the 
circle to each point and $\xi$ axis with the angles $\varphi_k$ giving the central phase of
the bin moving progressively along the histogram. 
The former angles can be computed from
   \begin{equation}
\sin~{\gamma_k} = (\eta_k - N/M) / R_k ~~~, ~~\cos~{\gamma_k} = (\xi_k - N/M) / R_k ~~~.
\label{eq:circularplot}
   \end{equation}

\noindent
and the relation between these two angles must be the straight line  
   \begin{equation}
\tilde{\gamma}_k = -2 \varphi_k + \psi = -2 (k-1/2) (360^\circ/M) + \psi
   \end{equation}

\noindent
where the factor 2 takes into account the second harmonic modulation.
We remark that the angle quadrant must be properly assigned to take into account
the two clockwise rounds described by the radius for increasing $k$.
For this reason the expected linear relation between $\gamma_k$ and $\varphi_k$ has
the slope equal to $-2$. 
The negative sign is due to the anticlockwise direction of data in the circular
plot; it can be inverted in computation without loss of information.

The resulting linear relations are shown in the right panel of Fig.~\ref{fig:anglin} 
for the original and shifted data sets.
The slope is very close to the expected value and also the intercepts
agree well with 0$^{\circ}$ (for the original data) and $-60^{\circ}$ (for the shifted data)
within the histogram bin width of $3^{\circ}.75$.

This calculation must be performed over $M$ values, corresponding to two complete rounds 
in the circle with increasing angles and consequently, for the bins in second half of the
histogram, a constant equal to 360$^{\circ}$ must be added when necessary to data and 
thus the two complete rounds correspond to 720$^{\circ}$. 
If $\langle \gamma \rangle$ would be computed over a single round ($M/2$ values) the 
constant 360$^{\circ}$ must be decreased to 180$^{\circ}$.
It is easy to verify that, considering the slope fixed at $-2$, the best estimate of $\psi$
is simply
   \begin{equation}
\psi = 360^{\circ} + \langle \gamma \rangle
\label{eq:psidef}
   \end{equation}

\noindent
as expected from a vertical shift of the data.

\subsection{Variance of the parameters of a circular plot}

It is interesting to study how the variance of the radius $R$ is related to the 
variance of the set $\{ n \}$.
   \begin{equation}
R_k^2 = (n_k - \langle n \rangle)^2 + (n_{k+M/8} - \langle n \rangle)^2
   \end{equation}

\noindent
and
   \begin{equation}
\langle R^2 \rangle = \frac{1}{M} \sum_{k=1}^{M} \bigg[(n_k - \langle n \rangle)^2 + 
(n_{k+M/8} - \langle n \rangle)^2 \bigg] = 2 \sigma_n^2 
   \end{equation}

\noindent
that for unpolarized data is equal the Poissonian variance $N/M$. One has also

   \begin{equation}
\Delta R_k = \sqrt{(n_k - \langle n \rangle)^2 + (n_{k+M/8} - \langle n \rangle)^2} - \langle R \rangle
   \end{equation}

\noindent
and
   \begin{equation}
\sigma_R^2 = \langle (\Delta R)^2 \rangle  = \frac{1}{M} \sum_{k=1}^{M} [R_k^2 + \langle R \rangle^2
-2 \langle R \rangle R_k ] = \langle R^2 \rangle  - \langle R \rangle^2  =  2 \sigma_n^2 - \langle R \rangle^2 
   \end{equation}
\noindent
then
   \begin{equation}
\sigma_R^2 = 2(\sigma_n^2 - \frac{1}{2} \langle R \rangle^2)   ~~~~~~~~.
   \end{equation}

The standard deviation of the mean $\langle R \rangle$ then will be
   \begin{equation}
\sigma_{\langle R \rangle} = \frac{1}{\sqrt{M}} ~\sigma_R = \frac{1}{\sqrt{M}} \sqrt{2 \sigma_n^2 - \langle R \rangle^2} ~~~~~~~~.
   \end{equation}

For the polarization degree we have:
   \begin{equation}
\sigma_{p} = \sigma_{\langle R \rangle} / [\mu (N/M)] =  \frac{\sqrt{M}}{\mu N} \sqrt{2 \sigma_n^2 - \langle R \rangle^2} ~~~~~~~~\label{eq:sigmap}.
   \end{equation}

From Eq.~\ref{eq:psidef} we see that the standard deviation of the polarization angle $\sigma_{\psi}$
is the same of $\langle \gamma \rangle$, which can be computed by standard error propagation
taking into account the right covariance term
$\langle \gamma R \rangle - \langle \gamma \rangle \langle R \rangle$
because these values are used in the calculation of the sine or cosine functions in
Eq.~\ref{eq:circularplot}.

\subsection{The histogram bin number and the $S/N$ ratio}\label{sec:snrratio}

The $S/N$ ratio of the polarization measure is given by $ \langle R \rangle / \sigma_R $
that measures how large is the distance of the mean circle from the its centre in units 
of the standard deviation of $R$. 
The use of $\sigma_{\langle R \rangle}$ instead of $ \sigma_R $ is not correct because it
does not take into account the actual dispersion of $R$ values around its mean and will 
give a higher $S/N$ value. 

The $S/N$ ratio is highly affected by the choice of the histogram number of bins $M$.
A high $M$, in fact, reduces the number of events in each bin
   \begin{figure} [ht]
   \begin{center}
   \begin{tabular}{c} %% tabular useful for creating an array of images 
   \includegraphics[width=6cm,height=6.0cm]{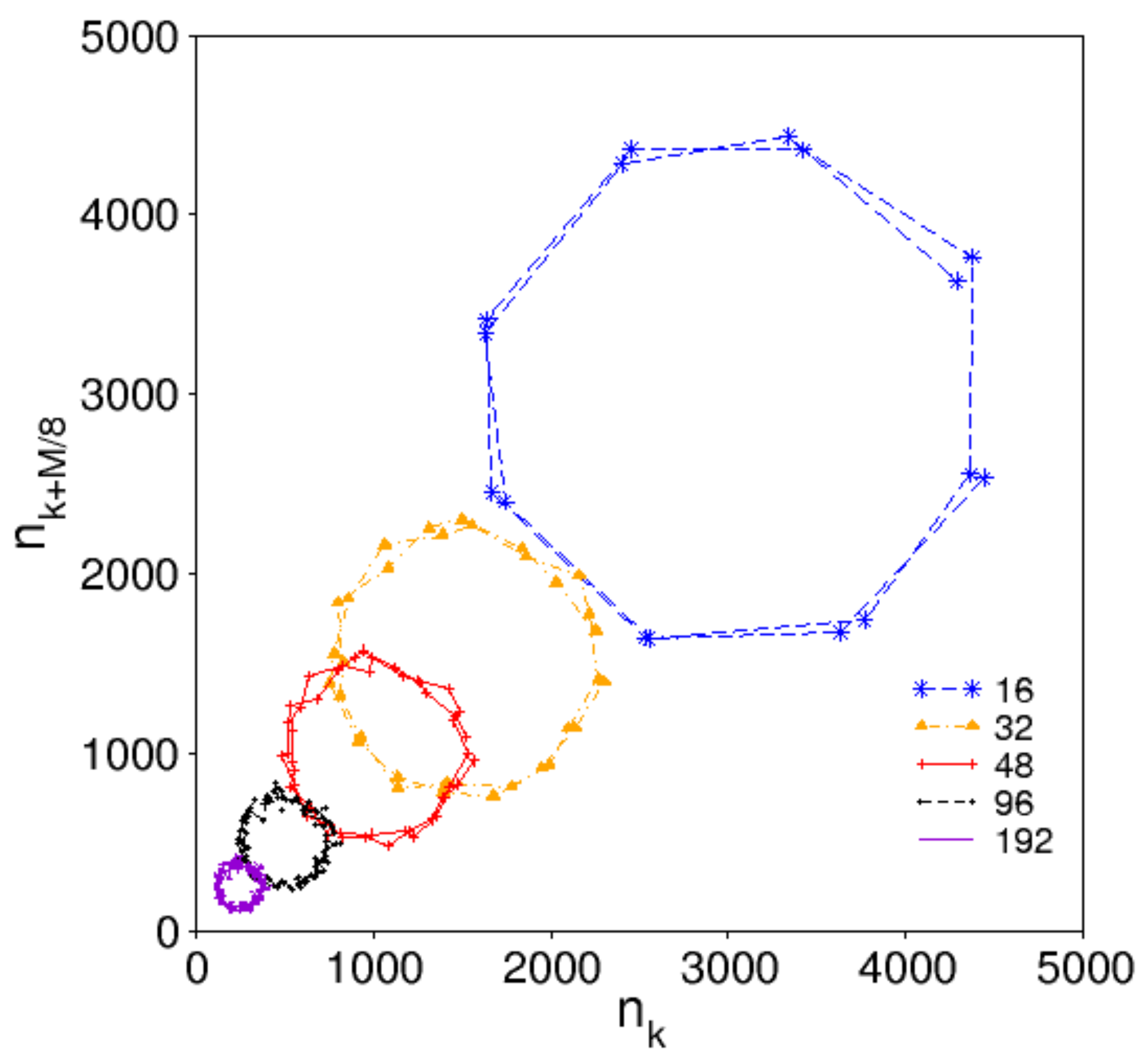}
   \includegraphics[width=6cm,height=5.9cm]{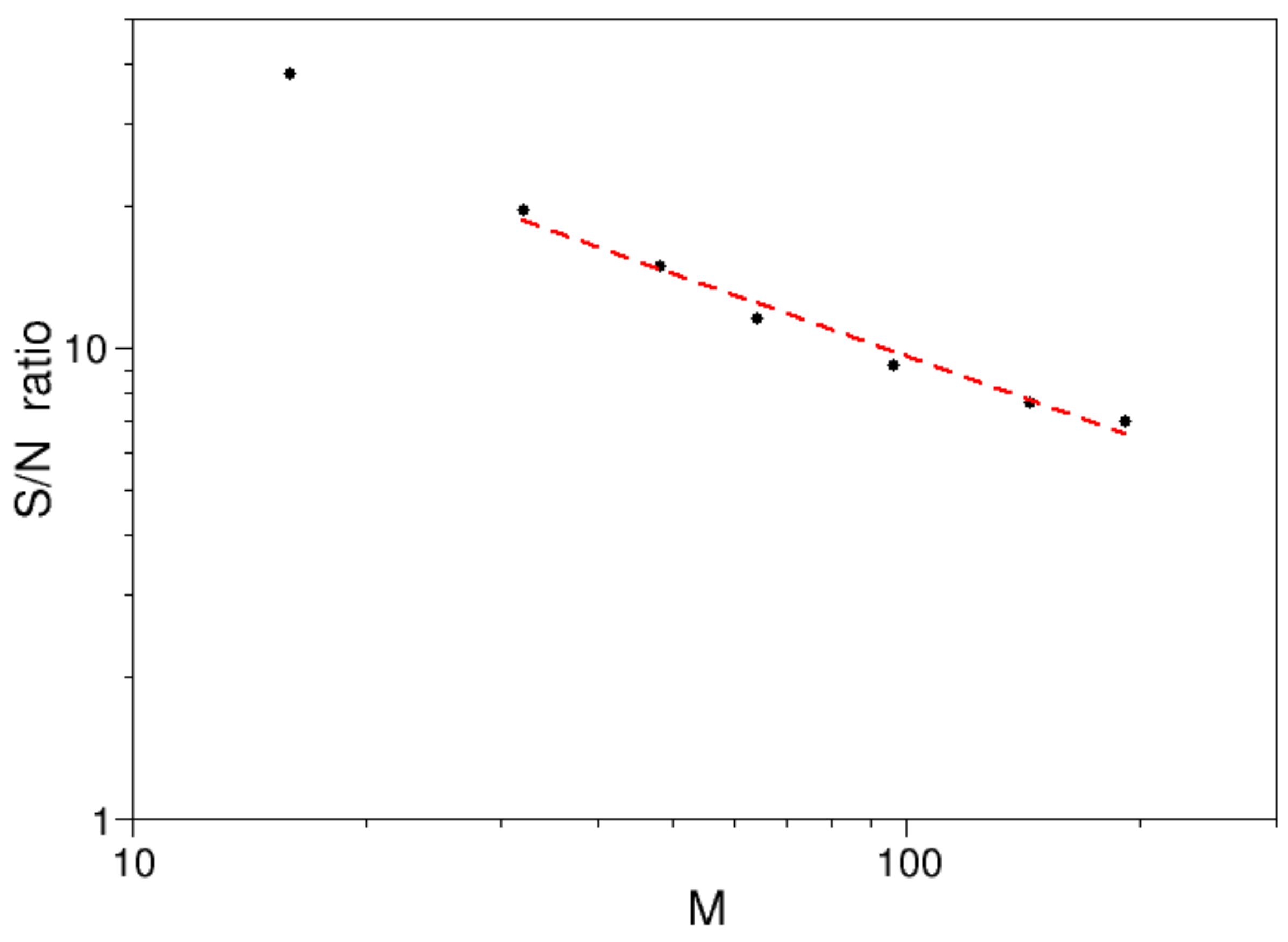}
   \end{tabular}
   \end{center}
   \caption[example] 
%>>>> use \label inside caption to get Fig. number with \ref{}
   { \label{fig:SNRplots} 
Left panel: circular plots of the experimental data of Fig.~\ref{fig:hist} with a 
different number of bins ranging from 16  to 192. 
Right panel: the $S/N$ of the same data set of the left panel for different number 
of bins $M$. 
The best fit with a power law from 32 to 192 bins is represented by the dashed red line. 
}
   \end{figure} 
and the statistical fluctuations on $n_k$ are approximately proportional to $\sqrt{M}$.
This is equivalent to a finer sampling and therefore to add higher frequency components 
in the Fourier spectrum: practically, the dispersion of the radius values increases, but 
the relative error on the mean value $\langle R \rangle$ remain stable because it is 
divided by $\sqrt{M}$.
Conversely, the effect of a low bin number is equivalent to apply a smoothing to the histogram, 
filtering the all the high frequency noise, with a consequent increase of the $S/N$ ratio.

To illustrate this effect we performed a test with some selected values of $M$, ranging from 
16 to 192, on the experimental data set in Fig.~\ref{fig:hist}.
The resulting circular plots are shown in the left panel of Fig.~\ref{fig:SNRplots}. 
The $S/N$ ratio evaluated by means of the previous formula changes from 38.3 and 20 for 16 and 32 
bins, respectively to 7.0 for 192 bins, while the relative error of the mean radius remains 
constant around 1\% as well as the accuracy of the estimates of the polarization degree and angle. 
The $S/N$ of the same data set is shown in the right panel Fig.~\ref{fig:SNRplots}. 
The best fit with a power law from 32 to 192 bins gives an exponent equal to $-0.58$, in a 
reasonable agreement with the expected $-0.5$
Note that in the left panel of Fig.~\ref{fig:SNRplots} the circle for $M = 16$ is reduced to 
an octagone because there are only 8 points for each round.
This is reasonably the lowest bin number to be used for detecting polarization, although the
use of 24 bins appears a right trade-off between an optimal $S/N$ ratio and well defined
circular plot.
For some instruments, as in the case of scattering polarimeters, the maximum bin number 
is fixed by the number of detectors and the use of 24 or 16 bins could not be admissible.
This problem should be considered in the design of the instrument.

\section{Numerical simulations}

\begin{figure} [ht]
   \begin{center}
   \begin{tabular}{c} %% tabular useful for creating an array of images 
   \includegraphics[height=7.8cm]{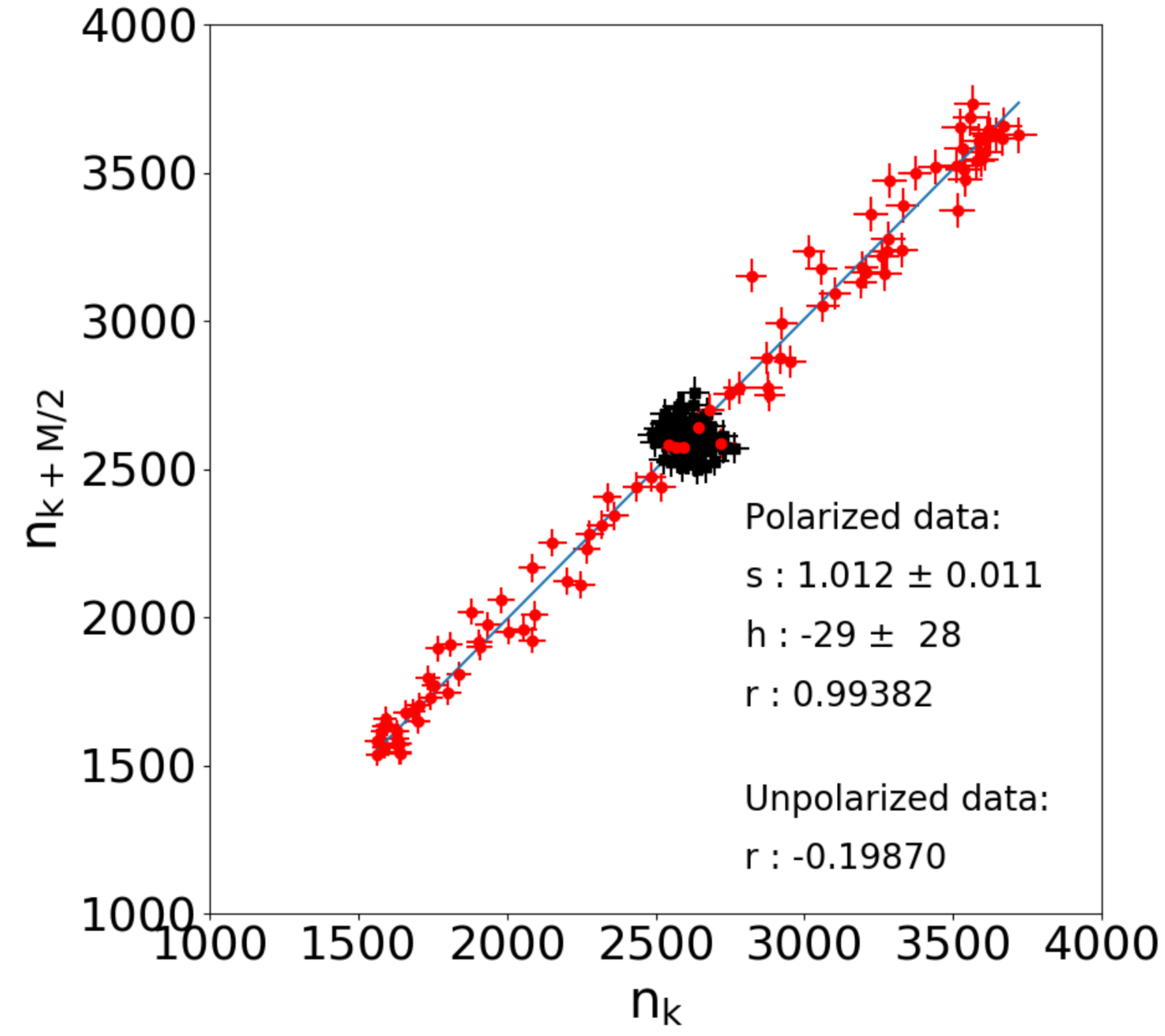} \\
   \includegraphics[height=8cm]{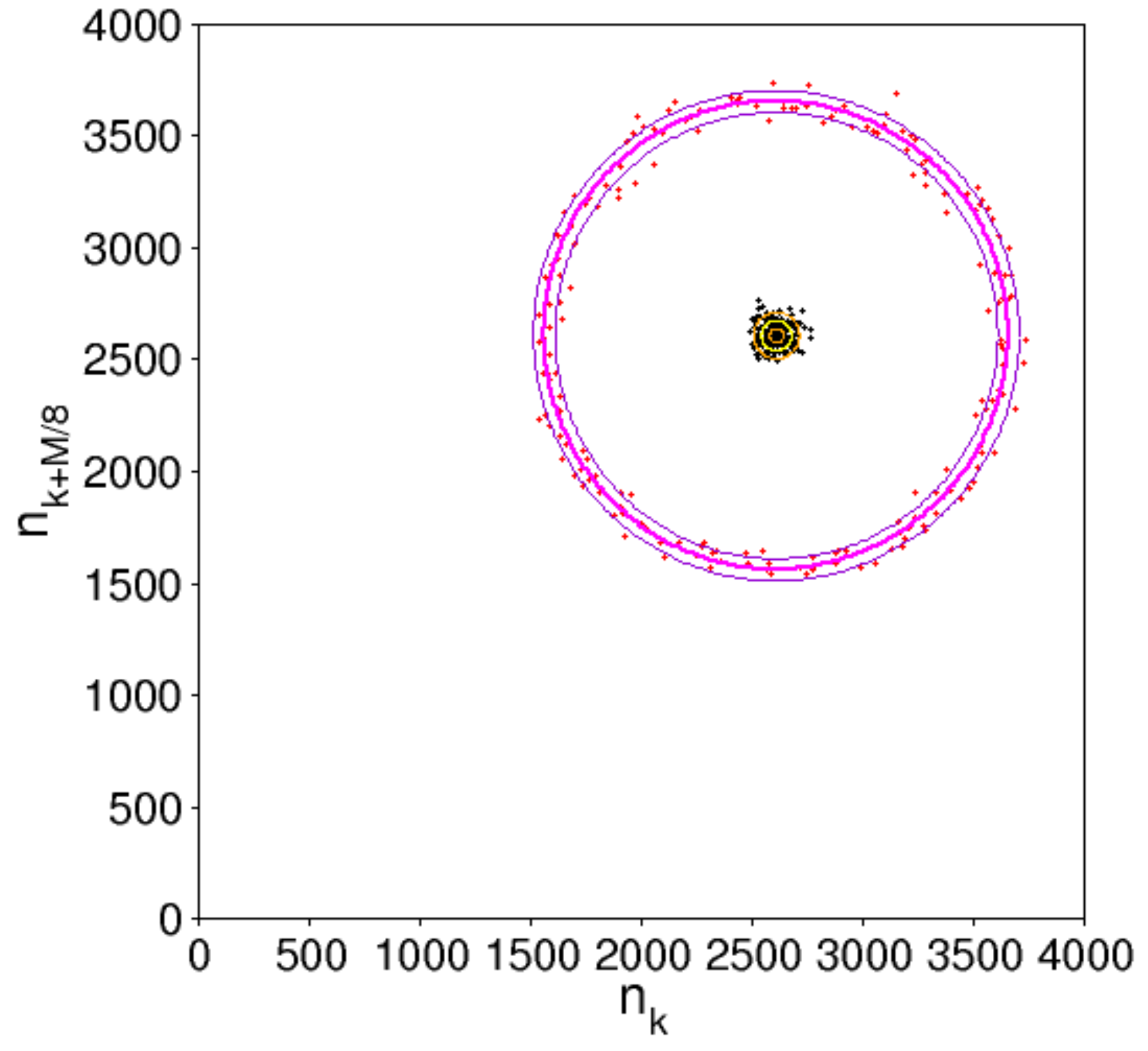}
   \end{tabular}
   \end{center}
   \caption[example] 
%>>>> use \label inside caption to get Fig. number with \ref{}
   { \label{fig:plotspolnonpol} 
Scatter plots of the linear (upper panel) and circular (lower panel) correlation for 
the same simulated data collected in 192 bin of the angular distribution. 
The filled squared points represent an unpolarized signal, while the filled circle 
points represent a signal $40\%$ polarized.
The solid line in the upper panel is the best fit of the polarized data.
The solid and dashed circles in the lower panel have radii equal to the average 
and $\pm 1\sigma$ outer and inner circles, respectively.
}
   \end{figure}

We used numerical simulations of polarization angular distributions, assuming that 
fluctuation are due to a Poissonian statistics, for investigating 
some statistical properties of the proposed tools.
Fig.~\ref{fig:plotspolnonpol} reports the same plots of Fig.~\ref{fig:plotscirclelinear} 
but for simulated data for an unpolarized (black points) and 40\% polarized signals 
with histograms of 192 angular bins for a total number of $5\times 10^5$ events.
Points of the former do not track any line (upper panel) or circle (lower panel) and 
appear randomly distributed around the mean value of counts. 
\clearpage

\clearpage

The capability of our tools for measuring the polarization is also shown by the linear
correlation plot in the upper panel of Fig.~\ref{fig:plotspolnonpol}: here the unpolarized data 
have a correlation coefficient has the low value $r = -0.1987$, while in the case of the 
polarized data it results $r = 0.9938$, and therefore only the $\approx$1.2\% of the 
total variance is due to statistical noise.
Note that in the circular plot the mean value of the radial distance between data points
and the centroid is always different from zero; in Fig.~\ref{fig:plotspolnonpol}  
it defines the solid circle, that in this case is 67.3, but this value is lower than 2 
standard deviation (outer and inner solid circles corresponding to 1$\sigma$), indicating 
that the resulting polarization degree is not statistically significant.

%\noindent
In the case of 40\% polarized histogram the mean radius of the circular distribution is $\langle R \rangle=1046.45$ with a standard deviation of $\sigma_R=48.72$. Therefore the S/N is more than 21 standard deviations.
The estimated value of the polarization degree is 0.4010$\pm$0.0013 (see Eq.~\ref{eq.polarizationR} and Eq.~\ref{eq:sigmap}), in a fully agreement with the simulation input parameter and coincident
with the sinusoidal best fit, as expected.

\subsection{Distribution functions of $R_k$}

It is possible to perform numerical simulations to obtain the histograms of the radial 
distances $R_k$:
in Fig.~\ref{fig:histsricerayl}  we present three~cases obtained from 192 bin histograms 
of an unpolarized signal, a 50$\%$ polarized signal and a 100$\%$ polarized signal. 
As apparent in Fig.~\ref{fig:plotspolnonpol} the data points are distributed with a 
random orientation with respect to the centroid of their coordinates, given by the 
mean value of the counts; therefore their radial distances follows the well known Rayleigh distribution:
   \begin{equation}
  f(R) = \frac{R}{\sigma^2}~e^{-R^2/2\sigma^2}   ~~~,\label{eq:Rayleigh}
   \end{equation}

\noindent
This formula is normalized to the unity and its mean value and variance are equal to 
$\sigma~\sqrt{\pi/2}$ and $ (2-\pi/2) \sigma^2$, respectively.

For a polarized signal the points spread around the centre located in the $\xi,\eta$ plane
at the position [$\langle n \rangle, \langle n \rangle$]; their distance to this centre 
changes because the fluctuations of counts in the phase histogram following a Poisson's
statistics that is very well approximated by a Gaussian law.
We stress that in principle the {\it thickness} of the point annulus is not uniform
because the Poissonian fluctuations are given by the square root of $n_k$ and therefore
their amplitude changes with the bin counts: we expect then a thickness larger in the
upper-right sector and narrower in the lower-left sector.
This effect can be easily estimated considering that the highest and lowest counts 
for a polarized signal with a resulting fractional modulation amplitude 
$a = A/\langle n \rangle$ are given by $\langle n \rangle (1 + a)$ and
$\langle n \rangle (1 - a)$, respectively; thus the difference of their standard
deviations are

   \begin{equation}
  \Delta \sigma = \sqrt{\langle n \rangle} (\sqrt{1 + a} - \sqrt{1 - a}) \approx  a \sqrt{\langle n \rangle} ~~~,
   \end{equation}

\noindent
thus, for typical modulation amplitudes $a < 0.3$  the change in the thickness of the
annulus with respect to the mean can be considered rather small.
Assuming that the annulus thickness is nearly constant, we approximated
the distribution of radius values with the slightly asymmetric law
given \citet{Plaszczynski2014}:

   \begin{figure} [ht]
   \begin{center}
%   \begin{tabular}{c} %% tabular useful for creating an array of images 
   \includegraphics[height=9.5cm,angle=0]{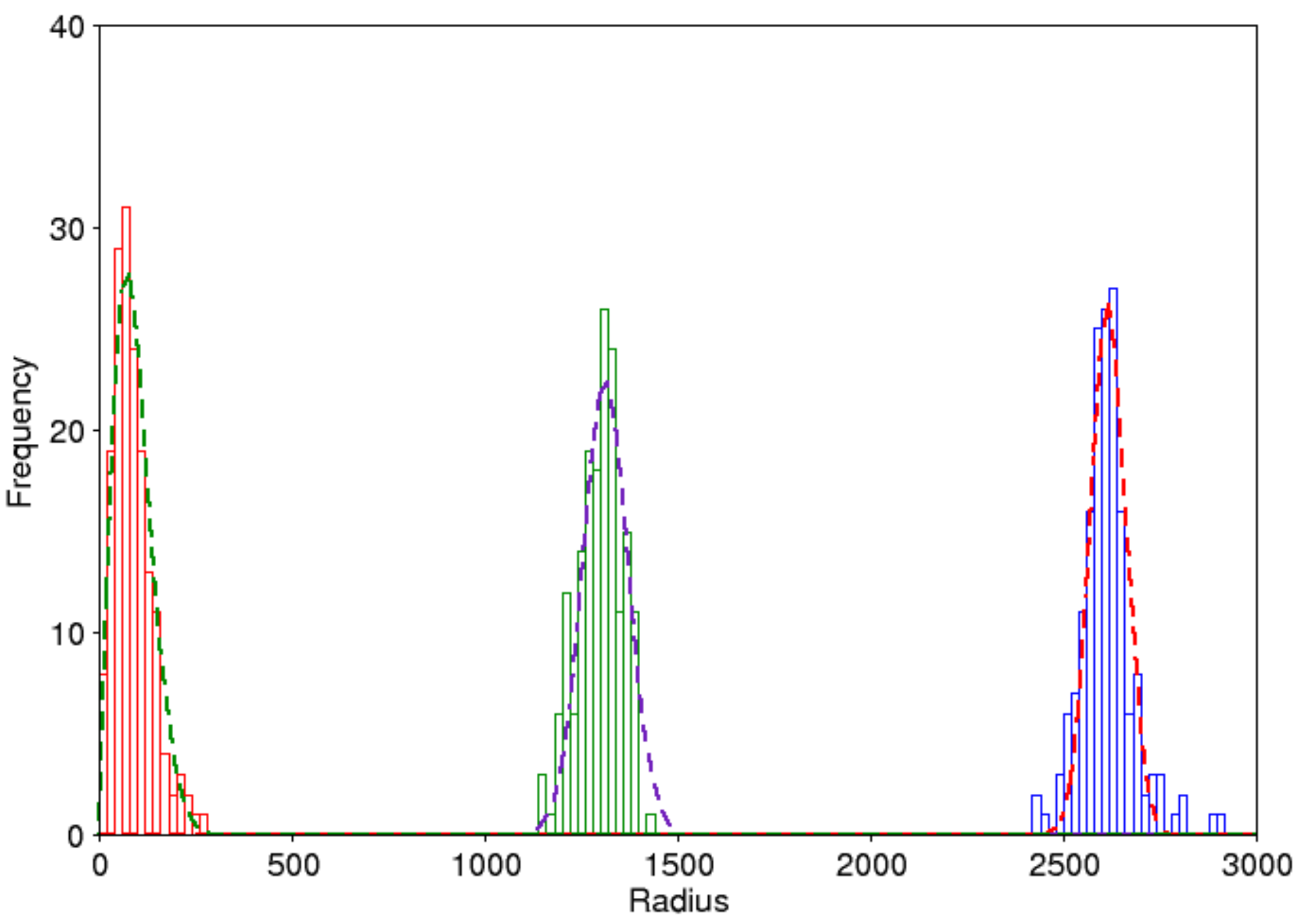} 
   \end{center}
   \caption[example] 
%>>>> use \label inside caption to get Fig. number with \ref{}
   { \label{fig:histsricerayl} 
Histograms of the radial values for the three simulated data data with different 
polarizations.
The histogram of the unpolarized signal (on the left side) appears
well described by a Rayleigh function given by the Eq.~\ref{eq:Rayleigh} (dashed line). 
The polarized signals at 50$\%$ (histogram in the center) and 100$\%$ (histogram on 
the right side) are described by the Plaszczynski et al. function as expressed in Eq.~\ref{eq:RiceApprox1} (fitted dashed lines).
}
   \end{figure} 

   \begin{equation}
  f(R) \approx \sqrt{\frac{R}{2 \pi \langle R \rangle}}~\frac{1}{\sigma}~\exp\{-(R - \langle R \rangle)^2/2\sigma^2\} ~~~.\label{eq:RiceApprox1}
   \end{equation}

In Fig.~\ref{fig:histsricerayl} is shown the radial distributions of the data resulting from 
simulations. 
The histogram of an unpolarized signal (on the left side) appears well described by a 
Rayleigh function given by the Eq.~\ref{eq:Rayleigh} (dashed line). 
The polarized signals at 50\% (histogram in the center) and 100\% (histogram on 
the right side) are described by the Plaszczynski et al. function as expressed in 
Eq.~\ref{eq:RiceApprox1} (dashed lines).
As a simple criterion Eq. 17 should be preferred to Eq. 19 when 
$\sigma \approx  \langle R \rangle$.
From these approximations one can compute the moments of the distributions useful for
statistical investigations.

\section{The Stokes' parameters}

The most usual method to estimate the polarization parameters is to compute the
Stokes' parameters, that in the case of a linear polarization measured by means of an
angular histogram of counts are (see, for instance, \citet{Kislat2015}):

   \begin{equation}
 Q = \sum_{k=1}^{M} n_k ~\cos(2 \varphi_k) ~~,~  U  = \sum_{k=1}^{M} n_k ~\sin(2 \varphi_k) ~~,~ I = \sum_{k=1}^{M} n_k~~
 \label{eq:stokes}
   \end{equation}
\noindent  from which one obtains
   \begin{equation}
 p = \frac{2}{\mu} \frac{\sqrt{Q^2 + U^2}}{I}  ~~~, ~~~ \psi = \frac{1}{2}\arctan(U/Q)~~~.\label{eq:polstokes}
   \end{equation}

\noindent

Of course, when applied to the previously considered data sets the results turns out 
fully consistent with other estimates: for instance, in the case of the simulated 
data of Fig.~\ref{fig:plotspolnonpol}, we found again $p = 0.401$, indicating that 
our circular plot method that is bias free.

It is clear from the scalar products of the first two formulae of Eq.~\ref{eq:stokes}  
that they are the second harmonic cosine and sine components of the Fourier series.
It is therefore very important to properly take into account the amplitude 
distribution over all the other frequencies due to noise fluctuations to 
evaluate the correct significance of the measure.
The estimate of the statistical uncertainties of the polarization parameters 
requires an accurate treatment as demonstrated in the paper by \citet{Kislat2015}. 

To compare the performances of these two methods we produced a set of simulated 
histograms with polarizion degrees varying from 100\% down to  1\% with an 
instrument modulation factor $\mu=0.5$ and a total number of events $N = 10^6$ 
distributed in 192 phase bins. 
A Poissonian noise was added to the counts in each bins with a variance equal 
to $n_k$.
   \begin{figure} [ht]
   \begin{center}
%   \begin{tabular}{c} %% tabular useful for creating an array of images 
   \includegraphics[height=6.0cm,angle=0]{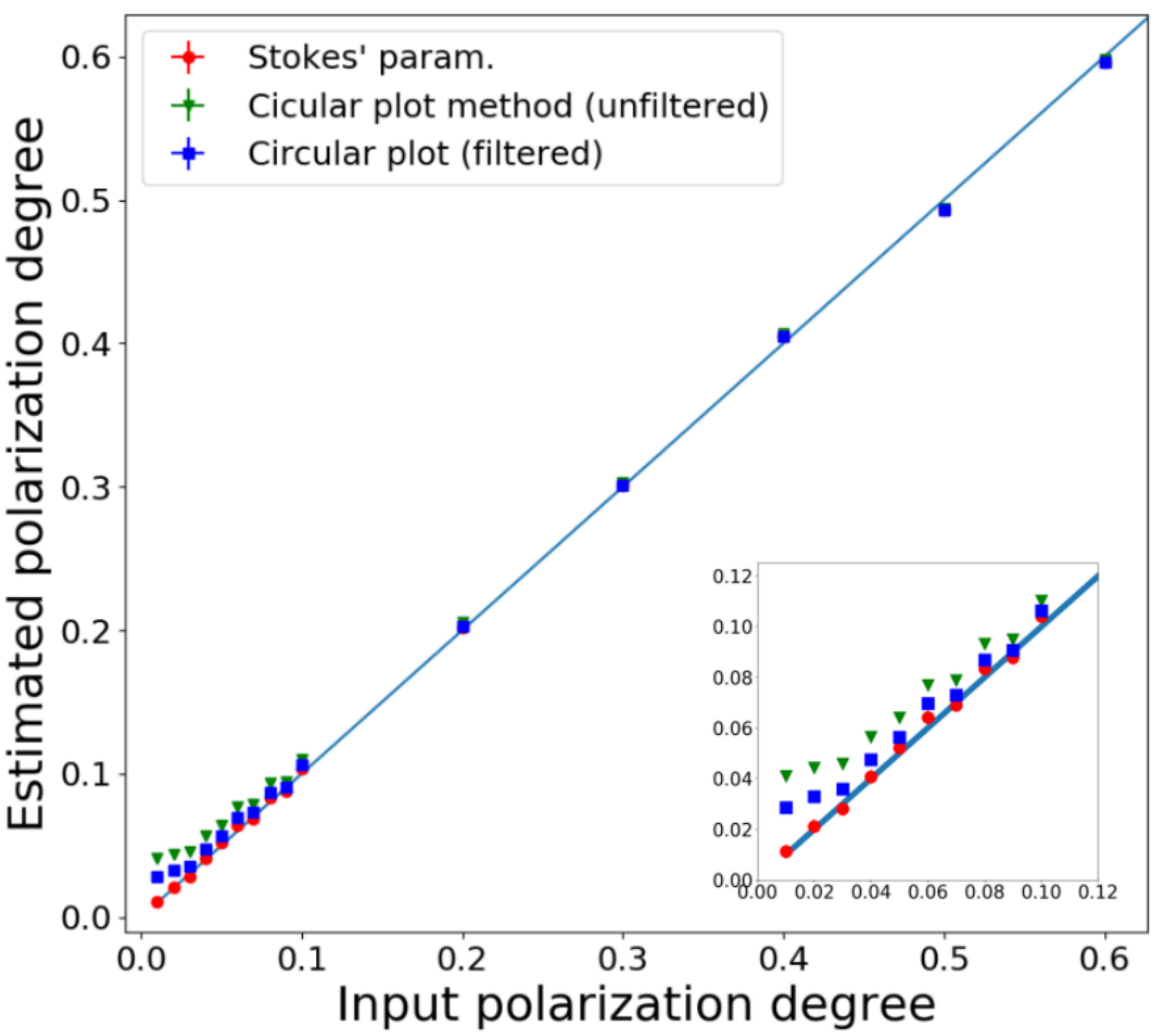} 
   \includegraphics[height=6.0cm,angle=0]{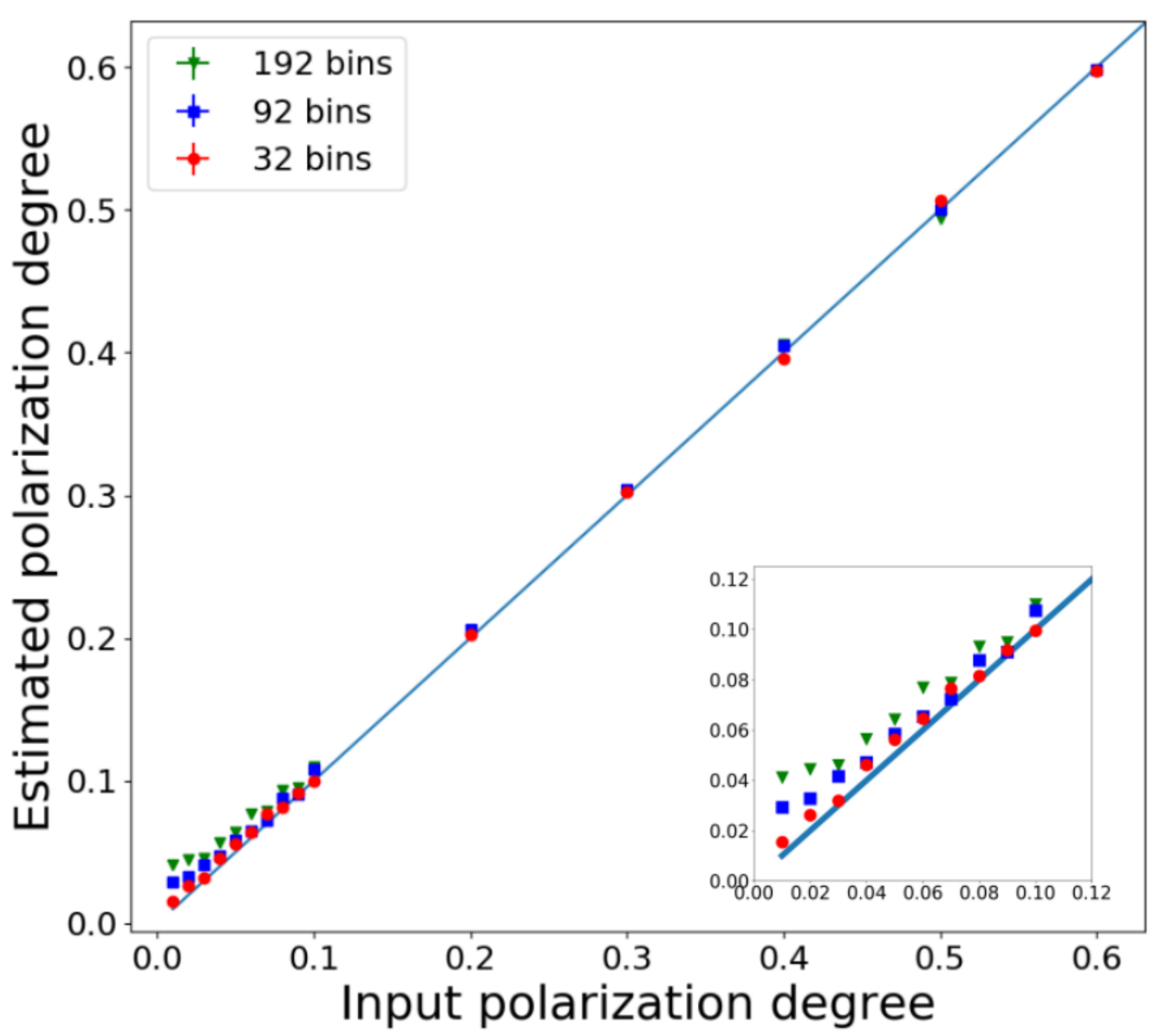} 
   \end{center}
   \caption[example] 
%>>>> use \label inside caption to get Fig. number with \ref{}
   { \label{fig:stokescircconfr} 
Left panel: comparison between polarization estimated obtained by means of 
the Stokes' parameters (filled circles) vs the circular plot method (unfiltered: 
downward triangles, filtered: filled squared points). 
Right panel: the comparison between polarization estimated obtained by applying 
the unfiltered circular correlation method with a different binning of the modulation 
histograms. 
The green downward triangles in two panels refer to the same data (unfiltered 
circular correlation metho) applied to a modulation histogram with 192 bins.
}
   \end{figure} 

In our method (circular plot) the statistical properties of data are entirely preserved.
This is clearly illustrated by the plot in the left panel of 
Fig.~\ref{fig:stokescircconfr}, where the results of the application of the two methods 
to simulated histograms are presented: the estimated $p$ values are coincident down to 
about 10\%; for lower values the polarization degree obtained by the Stokes' parameters 
(filled circles) remain very close to the expected ones, while those from the circular 
plot (downward triangles) are systematically higher approaching to a constant value 
of $\sim$4\%. 
For low polarization degrees the $R$ values practically follow a Rayleigh distribution 
whose mean value is highly depending on the $p/\sigma_p$ ratio.
After applying a noise reduction by means of simple running average smoothing over three 
consecutive histogram bins the resulting $p$ turns out to be closer to the expected values 
(filled squared points).
A stronger smoothing would further improve the estimates.

As discussed in Sect.~\ref{sec:snrratio}, the choice of a low bin number in the histogram is equivalent to a filtering.
We expect therefore that the results of the circular plots are more and more similar to those obtained by means of the Stokes' parameters when reducing the number of bins.
To verify this point we used the same simulation tool to produce histograms with $M$ ranging
from 32 to 192 and evaluated the polarization degree.
The results are plotted in the right panel of Fig.~\ref{fig:stokescircconfr} where
green downward triangles in both panels refer to the same data set (unfiltered circular 
correlation method) applied to a modulation histogram with 192 bins. 
It is clearly apparent that the use of the lowest $M$ value makes possible to detect 
polarizations as low as those obtained by means of the Stokes' parameters

This comparison shows also that the calculations of the Stokes' parameters based only on the 
amplitude of the second harmonic must be considered as a ``self-filtered'' method and therefore 
it does not take entirely into account the noise which affects all the other harmonics.
Thus, to evaluate the significance of the results one has to study the statistical 
properties of the histogram.
The use of the unfiltered circular plot contains all the information on the noise, 
including systematic deviations if present, and thus it provides a direct evaluation of 
the sensitivity reached in the measurements.

\section{Systematic effetcs and the correlation method}\label{systematics}
   \begin{figure} [h]
   \begin{center}
   \begin{tabular}{c} %% tabular useful for creating an array of images 
   \includegraphics[height=6.5cm]{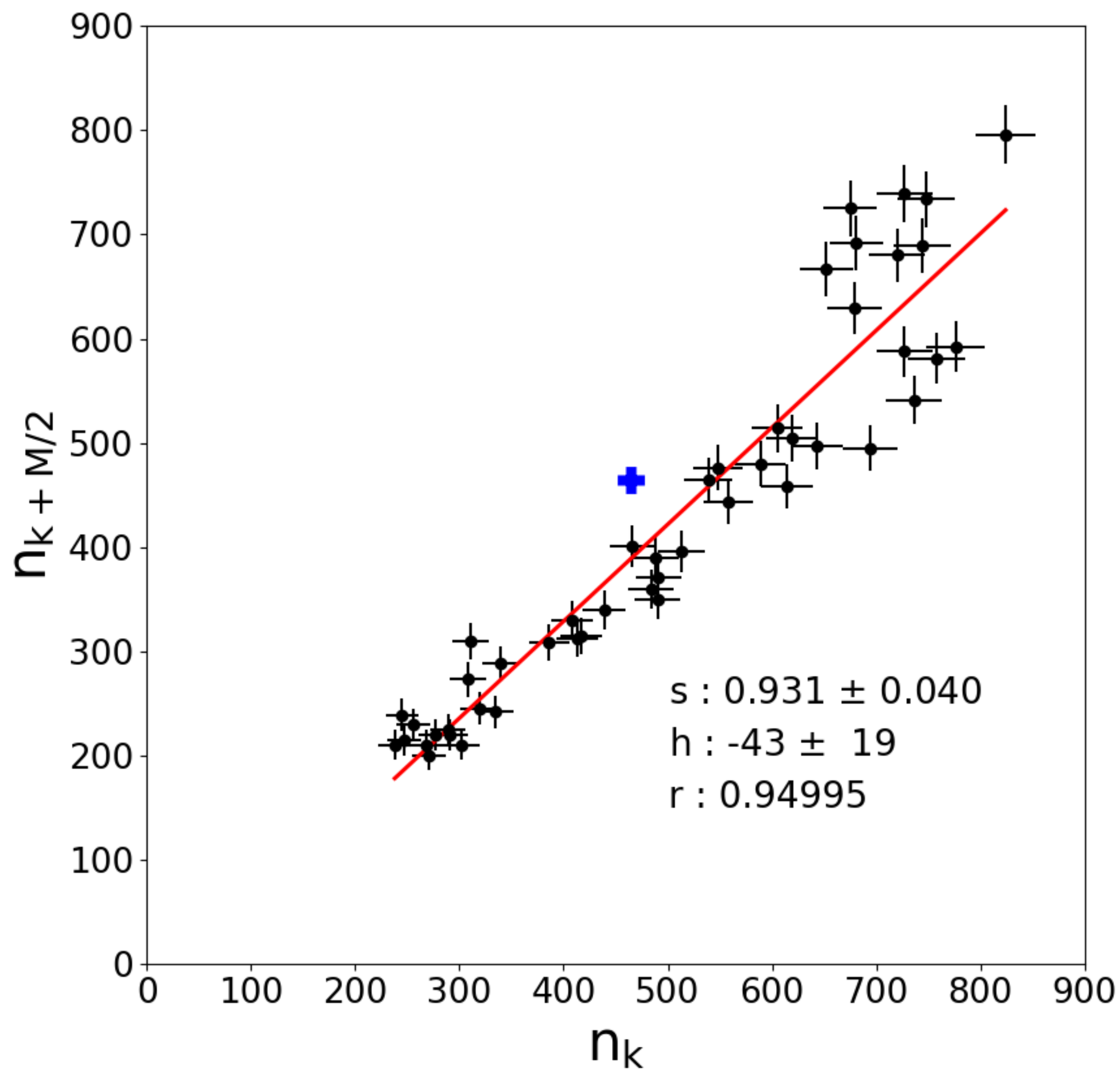} 
   \includegraphics[height=6.5cm]{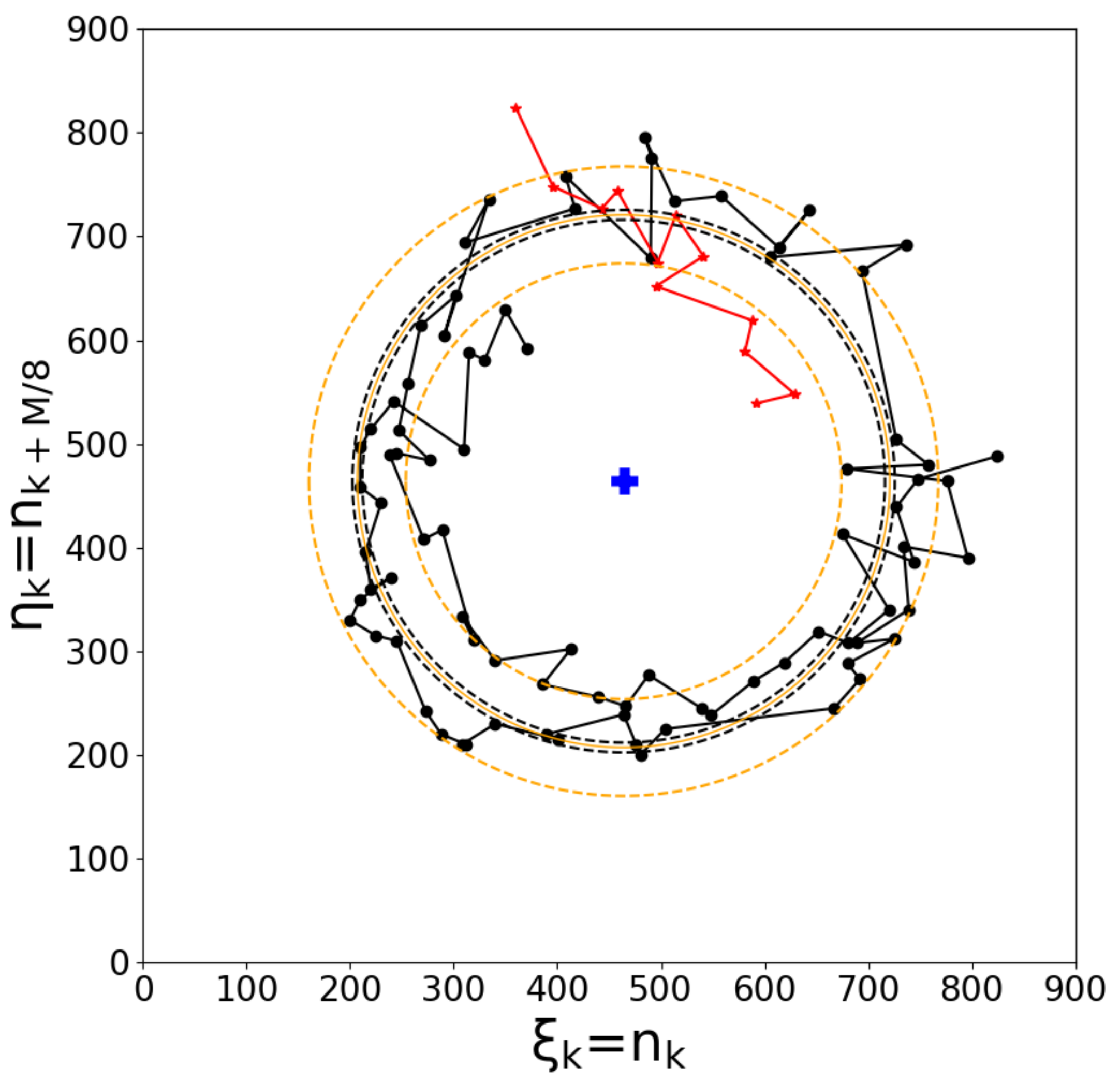} 
 \end{tabular}
   \end{center}
   \caption[example] 
%>>>> use \label inside caption to get Fig. number with \ref{}
   { \label{fig:systematics51020} 
Left panel: the linear correlation plots by assuming an event detection efficiency for 
angles larger than 210$^{\circ}$ decreased by 20\%.
The best fit values of the linear regression (orange line) and the correlation coefficient 
are reported. 
Right panel: the circular correlation plot in the same case. 
Note the splitting of the rings in two separated tracks.
}
   \end{figure}

The occurrence of systematics effects can affect the measurement and the use of 
correlation plots can help in their detection.
In scattering polarimeters, for example, a geometrical misalignment or a different 
efficiency of event detection in some angular channels can modify the azimuthal 
modulation of events in such a way that Eq.~\ref{eq:modulation2} does not describe 
properly any more the angular distribution of the signal modulation.  

We describe how a simple systematic effect modifies the correlation plots.
Let's assume a lowering of the gain that produces a decrease of detected counts for angular directions higher than a certain value. 
Then Eq.~\ref{eq:modulation2} is modified as follows by step function:
\begin{equation}
\mathrm{n}(\varphi_k)= n(\varphi_k) \cdot g(\varphi_k)
 \label{eq:modulationstep}
\end{equation} 
where
\begin{equation}
g(\varphi_k) = 
\begin{cases} 
1, & \mbox{if  } \varphi_k \leq  210^{\circ} \\ 
\varepsilon, & \mbox{if  } \varphi_k >  210^{\circ}  
\end{cases}
\end{equation}

We consider here the case of a high gain difference ($\varepsilon = 0.8$) to make more
evident how the correlation plots are modified.
The effects on the linear (left panel) and circular (right panel) correlation distribution 
are reported in Fig.~\ref{fig:systematics51020} which must be compared with those in 
Fig.~\ref{fig:plotscirclelinear}. 
In the linear correlation plot three changes are apparent: 
$i$) the expected mean value of counts per bin ($N/M$) is now different from the mean value in the first and second half of the phase histogram and therefore it is far from the regression line;
$ii$) the dispersion of data points with respect to the best fit is highly not uniform, 
being the points in the upper section much more scattered than in the lower one;
$iii$) the regression parameters are different from the expected ones, in particular
the slope is reduced to ($0.931\pm 0.040$) and the intercept $h$ is ($-0.43 \pm 19$) and it differs from the expected value by about 2 standard deviations. 

The circular plot of the same data shows a well clear separation of two branches, 
particularly apparent in the low count rate sector where the outer branch is due
to the lower count statistics on the bins after the fixed angular direction.
The same effect is present in the circular completion of the plot (red points).
We see, therefore, that even a simple inspection of these correlation plots can
be useful for highlighting possible deviations from the expected behaviour.

\section{Conclusion}

We proposed two simple tools for measuring the polarization parameters based on 
correlations between pairs of bins in the angular distribution having assigned 
shifts. 

We studied by means of laboratory data and numerical simulations how the results 
of the circular plot method depend upon the histogram binning.
In particular, we showed that in the correlation method the noise components at 
high frequencies is reduced by using a low number of bins in the modulation  
histogram as well as by applying a smoothing. 
No particular assumptions were made on the origin of the noise (for instance, 
if it is due entirely to Poissonian fluctuations or if there are other systematic 
contributions) for computing the $S/N$ ratio and variance of the estimated 
parameters' values. 
The estimate of the polarization degree and its significance is independent of 
the angle, that follows by a simple linear fit.
The results were found fully consistent with those obtained by other methods, 
as the evaluation of Stokes' parameters.
Moreover, the correlation plots are direct and efficent visual tools for fast 
detection of a polarized signal and the occurrence of systematic deviations.

Finally, the statistical properties of the circular distribution can be
derived in a straightforward way: for instance, for an unpolarized signal the
radial distribution of points in the circular plot follows the Rayleigh distribution
because their distances from the centre are those expected by a random gaussian 
process, while the one proposed by \cite{Plaszczynski2014} appears rather well 
appropriate distribution when there is a clear annular pattern due to the
presence of a significant polarization degree.

This method can be eventually included in the quick look data analysis of scientific 
data of XIPE \citep{Soffitta2016}.
Moreover, these tools are well suited for the analysis of data obtained by scattering 
polarimeters where the absorber is segmented into several scintillator elements (see, 
for instance, the COMPASS project by \citet{delMonte2016}) and the angular distribution 
is directly obtained as an histogram.
~\\

\section*{Acknowledgements}       
 
We are grateful to the referee K. Jahoda for his careful revision and germinating comments,
that contribute to make clearer our text.
This work was performed as an activity for the development of the XIPE (X-ray Imaging 
Polarimetry Explorer), IXPE (Imaging X-ray Polarimetry Explorer) and COMPASS (COMpton Polarimeter with Avalanche Silicon readout) 
projects.
XIPE is one of the three mission candidates to the M4 launch opportunity of the European 
Space Agency (ESA). The Italian participation to the XIPE study is funded by the Agreement 
ASI-INAF n. 2015-034-R.0. IXPE is an Explorer mission approved by NASA. Italian contribution to IXPE mission is supported by the Italian Space Agency through agreement ASI-INAF n.2017-12-H.0 and ASI-INFN agreement n.2017-13-H.0. COMPASS is funded under grant TECNO INAF 2014.

\section*{References}  

\bibliographystyle{elsarticle-harv}{}
%\bibliography{References.bib}

% References
%\bibliography{report} % bibliography data in report.bib
%\bibliographystyle{spiebib} % makes bibtex use spiebib.bst

\end{document}